\documentclass[prd,aps,twocolumn,showpacs,preprintnumbers,amsmath,amssymb,superscriptaddress,nofootinbib]{revtex4-1}
\usepackage[dvips]{graphicx}
\usepackage{epsf}
\usepackage{amsmath}
\usepackage{amssymb}
\usepackage{amsfonts}
\usepackage{times}
\usepackage[usenames]{color}
\usepackage[normalem]{ulem}
\usepackage[T1]{fontenc}
\usepackage[dvipsnames]{xcolor}

\usepackage{graphicx}% Include figure files
\usepackage{dcolumn}% Align table columns on decimal point
\usepackage{bm}% bold math

\begin{document}

\title{Revisiting the matter power spectra in $f(R)$ gravity}

\author{Jian-hua He}
\email[Email address: ]{jianhua.he@brera.inaf.it}
\affiliation{INAF-Observatorio Astronomico, di Brera, Via Emilio Bianchi, 46, I-23807, Merate (LC), Italy}

\author{Baojiu Li}
\affiliation{Institute for Computational Cosmology, Department of Physics, Durham University, Durham DH1 3LE, UK}

\author{Y.P. Jing}
\affiliation{Department of Physics and Astronomy, Shanghai Jiao Tong University, Shanghai 200240, China}

\pacs{98.80.-k,04.50.Kd}

\begin{abstract}
In this paper, we study the non-linear matter power spectrum in a specific family of $f(R)$ models that can reproduce the $\Lambda$CDM background expansion history, using high resolution $N$-body simulations based on the {\sc ecosmog} code. We measure the matter power spectrum in the range of $0.05h{\rm Mpc}^{-1}<k<10h{\rm Mpc}^{-1}$ from simulations for our $f(R)$ models and give theoretical explanations to their behaviour and evolution patterns. We also examine the chameleon mechanism for our models and find that it works throughout the cosmic history in dense regions, for our $f(R)$ models with $|f_{R0}|<10^{-4}$. On the other hand, for models with $|f_{R0}|>10^{-3}$, we find no chameleon screening in dense regions at late times ($z<3$), which means that those models could be ruled out due to the factor-of-$1/3$ enhancement to the strength of Newtonian gravity. We also give the best-fit parameters for a generalised PPF fitting formula which works well for the models studied here.
\end{abstract}

\maketitle
\section{Introduction}

Conclusive observational evidences from supernovae luminosity distances \cite{1}, cosmic microwave background (CMB) \cite{WMAP} and baryonic acoustic oscillations (BAO) \cite{BAOm} indicate that our Universe is undergoing a phase of accelerated expansion. Understanding the nature of this cosmic acceleration is one of the greatest challenges in contemporary physics. Theoretically, the leading explanation to it is a cosmological constant in the context of General Relativity (GR). Despite its notable success in describing the current cosmological data sets, this standard paradigm suffers from several problems: the measured value of the cosmological constant is far smaller than the prediction of the quantum field theory and there is a coincidence problem as to why the energy densities of matter and the vacuum energy are of the same order today (see \cite{sean} for review). It is also possible to explain the acceleration as driven by a mysterious component called dark energy, which is some kind of dynamical fluid with negative and time-dependent equation of state $w(a)$. However, to understand the nature of the dynamical dark energy is even harder than that of the cosmological constant in fundamental physics.

On the other hand, modified gravity theories are proposed as a promising alternative at explaining the observed accelerating expansion of our Universe. The idea is that GR might not be accurate on cosmological scales, and that the Universe may obey a different law of gravity. One of the simplest attempts is the so-called $f(R)$ gravity, in which the Ricci curvature $R$ in the Einstein-Hilbert action of GR is replaced by an arbitrary function of $R$ in the Lagrangian \cite{fr}. This model introduces an extra scalar degree of freedom which enables it to reproduce the accelerating expansion history of the universe with any effective dark energy equation of state $w(a)$ \cite{solution}. However, any specifically designed $w(a)$ other than $w=-1$ is less interesting because it can hardly be well-motivated in fundamental physics given the fact that we are still lack of knowledge about the nature of dark energy at the moment, and the observations do seem to favour $w=-1$. Therefore, it is of particular interest to investigate the family of $f(R)$ models that can exactly reproduce the $\Lambda$CDM background expansion history. The motivation behind this is threefold.

First, this family of $f(R)$ models can only be distinguished from the standard $\Lambda$CDM model in the perturbed space time, and any deviations from the $\Lambda$CDM growth history are direct consequences of the extra degree of freedom. This family of models can be considered as an ideal benchmark for testing the existence of scalar degrees of freedom in general modified gravity theories.

Second, the Brans-Dicke theory \cite{Brans,Dicke} and general coupled dark energy models \cite{Hefr} in the Einstein frame are equivalent to $f(R)$ gravity in the Jordan frame through conformal transformations as long as the distribution of the scalar curvature $R$ is continuous. This equivalence is rigorous in mathematics \cite{Maeda,conT,Hefr} and can also be well explained in physics \cite{Hefr,Fujii,Magnano}. Therefore, $f(R)$ gravity is not simply a stand-alone gravity theory but an equivalent representation for a wide class of modified gravity theories which involve extra scalar degrees of freedom.

Third, this family of $f(R)$ models do have the well-defined Lagrangian formalism in the spatially flat universe \cite{frmodel}, which is valid for the whole expansion history of the universe from the past to the future. The model is no longer simply a phenomenological model. The field equations can be deduced from the fundamental principle of least action. Moreover, the model has only one more extra parameter than that of the $\Lambda$CDM model.

Because of the importance of this specific family of $f(R)$ models, in this paper, we will further investigate the impact of the extra scalar degree of freedom on the large-scale structure in both the linear and the non-linear regimes using $N$-body simulations. We will first review the linear power spectrum for a large portion of parameter space using a modified version of {\sc camb} code \cite{frlinear} and address the importance of the chameleon mechanism \cite{Mota,Khoury} for $f(R)$ gravity to evade local tests of gravity. Then we will implement a large suite of $N$-body simulations based on {\sc ecosmog} code \cite{ECOSMOG} to examine the non-linear effect on the matter power spectrum of our $f(R)$ model.

This paper is organized as follows: In section~\ref{model}, we describe the details and summarize the distinct features of our $f(R)$ model. In section~\ref{linear}, we review the linear power spectrum of the model using accurate numerical results. In section~\ref{nonlinear}, we examine the non-linear power spectrum using a large suite of $N$-body simulations, and discuss the chameleon effect in our model. In section~\ref{conclusions}, we summarize and conclude this work.

\section{$f(R)$ cosmology \label{model}}

We work with the 4-dimensional action
\begin{equation}
S=\frac{1}{2\kappa^2}\int d^4x\sqrt{-g}[R+f(R)]+\int
d^4x\mathcal{L}^{(m)}\label{action}\quad,
\end{equation}
where $\kappa^2=8\pi G$ with $G$ being Newton's constant, $g$ is the determinant of the metric $g_{\mu\nu}$, $\mathcal{L}^{(m)}$ is the Lagrangian density for matter fields and $f(R)$ is an arbitrary function of the Ricci scalar $R$ \cite{fr} (see \cite{frreview,review_Tsujikawa} for reviews). In this work, we choose $f(R)$
to have the form of the Gaussian hypergeometric function \cite{frmodel}
\begin{equation}
\begin{split}
f(R)&=-\varpi\left (\frac{\Lambda}{R-4\Lambda}\right )^{p_+-1}{_2F_1}\left[q_+,p_+-1;r_+;-\frac{\Lambda}{R-4\Lambda}\right ]\\
&-2\Lambda\quad,
\end{split}\label{fr}
\end{equation}
which can enable the $f(R)$ model to mimic the $\Lambda$CDM background in a spatially flat universe. The indices in the expression are given by \cite{frmodel}
\begin{eqnarray}
q_+=\frac{1+\sqrt{73}}{12}\nonumber,\quad r_+=1+\frac{\sqrt{73}}{6}\nonumber, \quad p_+=\frac{5+\sqrt{73}}{12}\nonumber,
\end{eqnarray}
and $\varpi$ is a constant.

Hence, our model has only one more extra parameter than that of the $\Lambda$CDM model. Mathematically, when $b>0$ and $c>0$, the hypergeometric function ${_2F_1}[a,b;c;z]$ can have the integral representation on the real axis
\begin{equation}
{_2F_1}[a,b;c;z]=\frac{\Gamma(c)}{\Gamma(b)\Gamma(c-b)}\int_0^{1}t^{b-1}(1-t)^{c-b-1}(1-zt)^{-a}dt\quad,\label{defhypergeometric}
\end{equation}
where $\Gamma$ is the Euler gamma function. ${_2F_1}[a,b;c;z]$, in this case, is a real function in the range of $-\infty<z<1$ and our model Eq.~\ref{fr} is well-defined for $R>4\Lambda$. Moreover, it is important to note that our model does not have singularity although it appears to be divergent at $R=4\Lambda$. $f(R)$ is actually finite at $R=4\Lambda$ because we can find that
\begin{equation}
\begin{split}
&\lim_{R\rightarrow4\Lambda}f(R)=-2\Lambda\\
&-\frac{\varpi4(-511+79\sqrt{73})\Gamma(2/3)\Gamma(-r_-)}{(-5+\sqrt{73})(-1+\sqrt{73})(7+\sqrt{73})\Gamma(-p_-)\Gamma(q_+)}\\
&\approx-2\Lambda-1.256\varpi
\end{split}\label{lim4lambda}\quad,
\end{equation}
where
\begin{eqnarray}
r_-=1-\frac{\sqrt{73}}{6}\nonumber, \quad p_-=\frac{5-\sqrt{73}}{12}\nonumber.
\end{eqnarray}
When $R<4\Lambda$, Eq.~\ref{fr} becomes complex. Obviously, $R<4\Lambda$ is unphysical in our model.

For the background cosmology, we consider a homogenous and isotropic universe described by the flat Friedmann-Robertson-Walker (FRW) metric
\begin{equation}
ds^2=-dt^2+a^2d\mathbf{x}^2\quad.
\end{equation}
The modified Einstein
equation gives the modified Friedmann equation \cite{frmodel,frreview,review_Tsujikawa}
\begin{equation}
\begin{split}
&\frac{d^2f_{R}}{dx^2}+\left(\frac{1}{2}\frac{d\ln E}{dx}-1\right)\frac{df_{R}}{dx}+\frac{d\ln E}{dx}f_{R}\\
&=\frac{3(1+w)\Omega_d^0}{E}e^{-3\int_0^x(1+w)dx}\quad,\label{Gfield}
\end{split}
\end{equation}
where $f_R(x)\equiv\frac{\partial f}{\partial R}$  and $w$ is the effective dark energy equation of state, and the effective Friedmann equation $E\equiv\frac{H^2}{H_0^2}$ can be written as
\begin{equation}
E(x)=\Omega_m^0e^{-3x}+\Omega_d^0e^{-3\int_0^x(1+w)dx},\quad x\equiv\rm{ln}(a).\label{Ex}
\end{equation}
where the current dark matter density $\Omega_m^0$ and effective dark energy density $\Omega_d^0$ are defined by
\begin{equation}
\begin{split}
\Omega_m^0&\equiv\frac{\kappa^2\rho_m^0}{3H_0^2},\\
\Omega_d^0&\equiv\frac{\kappa^2\rho_d^0}{3H_0^2}.\label{defination}
\end{split}
\end{equation}

The background expansion history of our $f(R)$ model can exactly mimic that of the $\Lambda$CDM paradigm from the matter dominated epoch to the future, which yields very simple expressions for the background evolution
\begin{equation}
\begin{split}
E(x)&=\Omega_m^0e^{-3x}+\Omega_d^0\quad,\\
R(x)&=[3\Omega_m^0e^{-3x}+12\Omega_d^0]H_0^2\quad,
\end{split}
\end{equation}
where $R$ is the scalar curvature.

The $f(R)$ cosmology differs from the standard $\Lambda$CDM cosmology by an additional scalar degree of freedom. As we shall see later, this scalar degree
of freedom plays an important  role in the perturbed space-time in  $f(R)$ gravity. In the background, the evolution of the scalar field $f_{R}$ is governed by Eq.~\ref{Gfield}. However, in our model, $f_{R}$ has an explicit expression which is the exact solution to Eq.~\ref{Gfield} with $w=-1$ \cite{frmodel}
\begin{equation}
\begin{split}
f_R(x)&=D(e^{3x})^{p_+}{_2F_1}\left[q_+,p_+;r_+;-e^{3x}\frac{\Omega_d^0}{\Omega_m^0}\right] \quad, \label{mG}
\end{split}
\end{equation}
where $D$ is a dimensionless quantity, and is related to the covariant parameter $\varpi$ in Eq.~\ref{fr} by
\begin{equation}
\begin{split}
\varpi&=D(R_0-4\Lambda)^{p_+}/(p_+-1)/\Lambda^{p_+-1}\\
      &=\frac{D}{p_+-1}\left(\frac{\Omega_m^0}{\Omega_d^0}\right)^{p_+}3\Omega_d^0 H_0^2\quad.\label{def_D}
\end{split}
\end{equation}
For more details about our model, we refer readers to \cite{frmodel}.

At early times, the universe is dominated by matter and the curvature is very high $R\gg 4\Lambda$. The hypergeometric function goes back to unity ${_2F_1}\sim1$. Thus, Eq.~\ref{fr} can reduce to
\begin{equation}
f(R)\sim -\varpi\left (\frac{\Lambda}{R}\right )^{p_+-1}\quad,
\end{equation}
which can exactly mimic the $\Lambda$CDM background in the matter dominated epoch. Moreover, for higher scalar curvature $R\rightarrow+\infty$, our model goes back to standard GR
\begin{equation}
\begin{split}
\lim_{R\rightarrow+\infty}f_{R}(R)&=0\quad.
\end{split}
\end{equation}

On the other hand, in the future limit ($x\rightarrow +\infty$) where the energy density of matter fields tends to be zero ($\rho_m \rightarrow 0$), the universe is almost empty and dominated only by vacuum. The scalar curvature $R$ goes as $R\rightarrow 12\Omega_d^0H^2_0=4\Lambda$ rather than zero. From Eq.~\ref{lim4lambda}, we can see clearly that Eq.~\ref{fr} is not divergent at $R=4\Lambda$ , which means that our model is able to describe the universe even in the extreme case of vacuum. Our model, therefore, is self-consistent and is valid throughout the cosmic history.

In summary, our model has the well-defined Lagrangian formalism.  The model is not merely a phenomenological one, and its field equations can be derived from the principle of least action. Our model has only one extra parameter compared with $\Lambda$CDM model, and it can exactly reproduce the $\Lambda$CDM background expansion history from the past to the future.
When $\varpi \neq 0$, the constant $\Lambda$ in Eq.~\ref{fr} cannot be explained as the energy density of the vacuum although it takes the same value as $\Lambda$ in the $\Lambda$CDM model and our model does not suffer the cosmological constant problem. Moreover, when $D<0$ and $|f_{R0}|<1$, our model satisfies:
\begin{enumerate}
\item $1+f_{R}>0$ for $R\geq R_0$, where $R_0$ is the Ricci scalar today.
\item $f_{RR}>0$ for $R\geq R_0$.
\item $R+f(R)\rightarrow R-2\Lambda $ for $R\geq R_0$.
\item Obviously, our model can achieve the late-time acceleration since it reproduces the $\Lambda$CDM background expansion history.
\end{enumerate}
Our model, therefore, meets the requirements for the viable metric $f(R)$ models as proposed in \cite{review_Tsujikawa}.

\section{The linear matter power spectra\label{linear}}

\begin{figure}
\includegraphics[width=3.6in,height=3.2in]{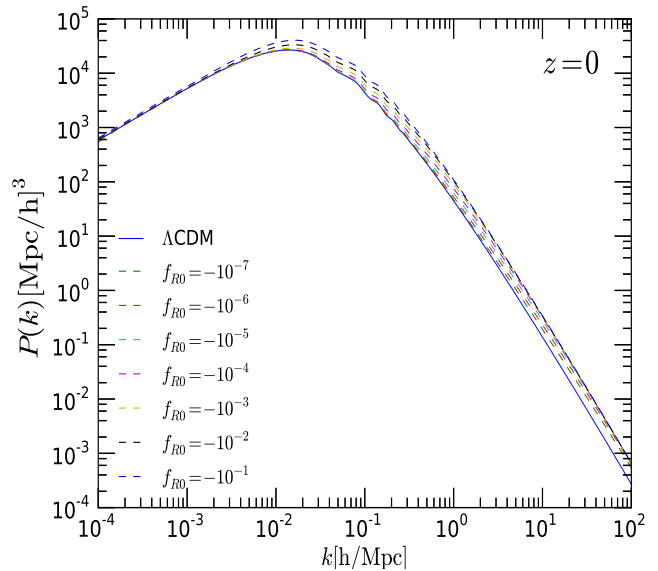}
\caption{The linear matter power spectrum for our $f(R)$ models.} \label{pklinear}
\end{figure}

\begin{figure}
\includegraphics[width=3.6in,height=3.2in]{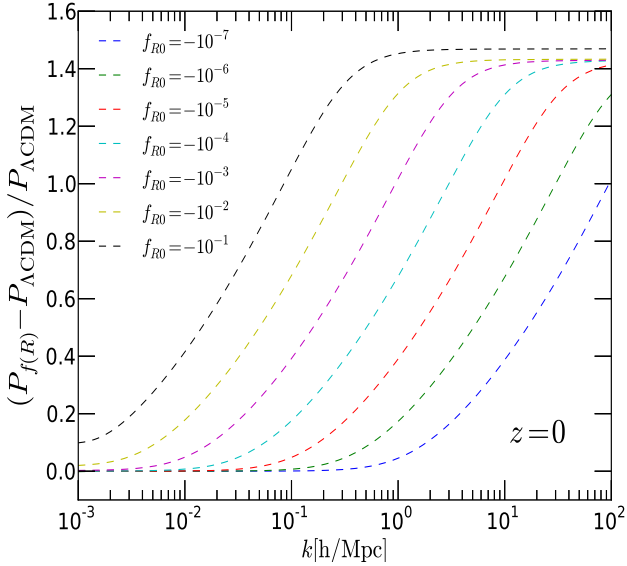}
\caption{The relative difference of the linear matter power
spectra between the $f(R)$ models and the $\Lambda$CDM model at $z = 0$.} \label{frpklin}
\end{figure}

In this work, we calculate the accurate linear matter power spectra  using our modified version of the CAMB code~\cite{CAMB} which solves the full linear perturbation equations in $f(R)$ gravity \cite{frlinear}. We set the initial conditions for the linear scalar field perturbations at $a=0.04839$ as $\delta f_{R}=0$ and $\delta f'_{R}=0$ where prime denotes to the derivative with respect to the conformal time, and assume the cosmological parameters as $\Omega_m^0=0.2814,\Omega_d^0=0.7186,h=0.697,n_s=0.962,\sigma_8=0.82$ throughout this work.
The numerical results are shown in Fig.~\ref{pklinear} and Fig.~\ref{frpklin}. In Fig.~\ref{pklinear}, we show the
linear matter power spectra for a large range of scales ($10^{-4}h{\rm  Mpc^{-1}}<k<10^{2}h{\rm Mpc^{-1}}$) and of the parameter $10^{-7}<-f_{R0}<10^{-2}$. In Fig.~\ref{frpklin}, we illustrate the fractional difference  between $f(R)$
gravity and general relativity in the matter power spectrum. In order to better explain our numerical results, we illustrate here with the aid of a simplified equation for the growth history of $f(R)$ gravity \cite{frreview}
\begin{equation}
\ddot{\delta}_m+2H\dot{\delta}_m-4\pi G_{\rm eff}\rho_m\delta_m=0\label{quasi}
\end{equation}
where $\delta_m$ is the density contrast for matter field,  dot denotes the derivative with respect to the cosmic time and the effective Newtonian constant $G_{\rm eff}$ is given by \cite{frreview}
\begin{equation}
G_{\rm eff}\equiv\frac{G}{1+f_R}\frac{4+3M^2a^2/k^2}{3(1+M^2a^2/k^2)}\quad,
\end{equation}
where
\begin{equation}
M^2=\frac{1}{3}\left(\frac{1+f_R}{f_{RR}}-R\right)\quad,
\end{equation}
is the mass squared for the scalar field. Eq.~\ref{quasi}, actually, can not give the accurate growth history for $f(R)$ models as pointed out in \cite{Dombriz}. However it does give the correct qualitative behaviors at some extreme cases. We use this simplified equation here only for illustrative purpose.

First, on very small scales $k>10h{\rm Mpc^{-1}}$,
the growth history becomes scale-independent regardless the types of $f(R)$ models. No matter how small we choose the parameter $|f_{R0}|$, there is a factor of $\frac{4}{3}$ enhancement in the effective Newtonian constant as $k$ trends to infinity,
\begin{equation}
\lim_{k \rightarrow +\infty}G_{\rm eff}=\frac{4G}{3(1+f_R)}\quad,
\end{equation}
and this is known as ``scalar-tensor'' \cite{frreview} or equivalently ``low-curvature'' \cite{HuI} regime. The curvature $\delta R$ is well suppressed and no longer tracks the matter density field (as it does in GR). The enhancement in the effective Newtonian constant $G_{\rm eff}$ could render $f(R)$ gravity models unable to pass the local tests. On the other hand, this enhancement would also increase the linear power of matter in $f(R)$ gravity on the smallest scales at present time compared to the $\Lambda$CDM model. As a result, the ratio $(P_{f(R)}-P_{\rm \Lambda CDM})/P_{\rm \Lambda CDM}$ in Fig.~\ref{frpklin} trends to be a constant on extreme small scales ($k>10h{\rm Mpc^{-1}}$) even for the smallest value of $|f_{R0}|=10^{-7}$ as chosen in our plots. Analytically, this can be understood as following: the solution of Eq.~\ref{quasi} for the growth history in $\Lambda$CDM model ($G_{\rm eff}=G$) is $\delta_m^2\propto t^{4/3}$ and, on extrame small scales $k>10h{\rm Mpc^{-1}}$, the solution for $f(R)$ gravity with $G_{\rm eff}=\frac{3}{4}G$ is $\delta_m^2\propto t^{(\sqrt{33}-1)/3}$ \cite{frreview}. The ratio of the matter power spectrum, therefore, is
\begin{equation}
\frac{P_{f(R)}}{P_{\rm \Lambda CDM}}\propto t^{(\sqrt{33}-5)/3}\quad,
\end{equation}
which is scale-independent and only depends on the initial conditions.

Second, in the small wave number $k$ limit ($M^2\geq k^2/a^2$), $f(R)$ gravity will become very close to GR as
\begin{equation}
\lim_{f_{R0}\rightarrow 0}G_{\rm eff}=\frac{G}{1+f_{R0}}\quad.
\end{equation}
This is known as ``general relativistic regime'' \cite{frreview} or equivalently ``hight-curvature regime'' \cite{HuI} where the curvature $\delta R$ is able to track the matter density field ($\delta R\sim\kappa^2\delta \rho$) even in the case that $\delta \rho$ is very small. However, the influence of the factor $1+f_{R0}$ could be prominent when the absolute value of $f_{R0}$ approaches unity. The amplitude of the power spectrum will be enhanced due to the factor of $\frac{1}{1+f_{R0}}$ (remember that $f_{R0}<0$), which is shown clearly in Fig.~\ref{frpklin}.

Third, the scale of the transition from the ``high-curvature regime'' to the ``low curvature regime'' can be characterized by the Compton wavelength which is defined by \cite{Song}
\begin{equation}\label{eq:compton}
B=\frac{f_{RR}}{1+f_R}\frac{dR}{dx}\frac{H}{\frac{dH}{dx}}\quad.
\end{equation}
In our model, we can find an analytical relation between $D$ and the Compton wavelength today $B_0\equiv B(a=1)$, as
\begin{widetext}
\begin{equation}
\begin{split}
B_0&=\frac{2D p_+}{(\Omega_m^0)^2\left\{1+D{_2F_1}\left[q_+,p_+;r_+;-\frac{\Omega_d^0}{\Omega_m^0}\right]\right\}}\times\left\{\frac{q_+}{r_+}\Omega_d^0{_2F_1}\left[q_++1,p_++1;r_++1;-\frac{\Omega_d^0}{\Omega_m^0}\right]-\Omega_m^0{_2F_1}\left[q_+,p_+;r_+;-\frac{\Omega_d^0}{\Omega_m^0}\right]\right\}.
\end{split}
\end{equation}
\end{widetext}
We can also find the relationship between $D$ and $f_{R0}$ as
\begin{equation}
\begin{split}
f_{R0}&=D\times{_2F_1}\left[q_+,p_+;r_+;-\frac{\Omega_d^0}{\Omega_m^0}\right]\quad\label{mG}.
\end{split}
\end{equation}
Thus the value of Compton wavelength $B_0$ is only determined by $f_{R0}$ if the background cosmology is fixed. The diminishing value of $|f_{R0}|$ will push the transition between different regimes toward smaller scales. For any given wave number $k$ or a certain scale we are interested in, smaller absolute value of $|f_{R0}|\rightarrow 0$ will enhance the mass squared ($M^2$) for the scalar field. $M^2$ is able to surpass the wave number $M^2\geq k^2/a^2$ and the effective Newtonian constant could go back to the ``general relativistic regime''
\begin{equation}
\lim_{D\rightarrow 0}G_{\rm eff}=\lim_{f_{R0}\rightarrow 0}G_{\rm eff}=G.
\end{equation}
This phenomena is consistent with our naive expectation that setting $\varpi=0$ in Eq.~\ref{fr} forces the model back to standard $\Lambda$CDM. Of course, this is only an extreme case which means that $\lim_{f_{R0}\rightarrow 0}B_0=0$ such that the transition happens on extremely small scales which is very close to zero.

In summary, according to linear theory, if $f_{R0}\neq0$, the $f(R)$ gravity model would always have the ``low-curvature solution'' on extreme small scales no matter how small $|f_{R0}|$ is.
The factor-of-$1/3$ enhancement to the strength of Newtonian gravity on small scales would make the $f(R)$ theory fail to pass the local test.  Fortunately, if the chameleon mechanism \cite{Mota,Khoury,li_barrow_2007,HuI,brax_2008} works efficiently the model could still follow the ``high-curvature'' solution in high-density regions at late times of the Universe and the ``low-curvature regime'' only appears in low-density regions on scales where the Compton condition is violated\cite{HuI}. This class of models could then pass local experimental constraints in high-density regions such as our solar system.

\section{Non-linear Power spectra\label{nonlinear}}

To study the non-linear power spectra, we carry out a large suite of $N$-body simulations, which are based on the {\sc ecosmog} code developed by \cite{ECOSMOG}. {\sc ecosmog} is a modified version of the mesh-based $N$-body code {\sc ramses} \cite{RAMSES}, which calculates the gravitational force by solving the Poisson equation on meshes
using a relaxation method to obtain the Newtonian potential and
then differencing the potential. {\sc ecosmog} is efficiently
parallelised and suitable to run simulations systematically.

In $N$-body simulations, at early times and in high density regions, we assume that $R\gg 4\Lambda$ and the hypergeometric function goes back to unity ${_2F_1}\sim1$. Eq.~\ref{fr} reduces to
\begin{equation}
f(R)\sim -\varpi\left (\frac{\Lambda}{R}\right )^{p_+-1}\quad.\label{matterfr}
\end{equation}
Although Eq.~\ref{matterfr} is much simpler than Eq.~\ref{fr}, we stress that the model it represents can exactly mimic the $\Lambda$CDM background in the matter dominated epoch no matter how large we choose the value of $\varpi$.

Taking the derivative of above equation and using Eq.~\ref{def_D}, we find that
\begin{equation}
f_R(R)\sim D\left(\frac{3\Omega_m^0H_0^2}{R}\right)^{p_+}<0, \quad D<0\quad.
\end{equation}
Inversely, we can obtain $R$ in terms of $f_R$
\begin{equation}
R=3\Omega_m^0H_0^2\left(\frac{D}{f_R}\right)^{\frac{1}{p_+}}\quad.\label{Rcur}
\end{equation}

\subsection{High-curvature and Low-curvature solutions}

In $f(R)$ gravity the structure formation is governed by the modified Poisson equation
\begin{equation}
\nabla^2\phi=\frac{16\pi G}{3}\delta \rho-\frac{\delta R}{6}\quad,\label{poissonfr}
\end{equation}
as well as the equation for the scalar field $f_{R}$ \cite{HuI}
\begin{equation}
\nabla^2\delta f_R=\frac{1}{3c^2}[\delta R - 8\pi G\delta \rho]\quad,\label{frpoisson}
\end{equation}
where $\phi$ represents the gravitational potential, \quad$\delta f_R=f_R(R)-f_R(\bar{R}),\quad\delta R= R-\bar{R},\quad\delta \rho=\rho-\bar{\rho}$. The overbar denotes the background quantities, and $\nabla$ is the gradient operator with respect to the proper distance.
Inserting Eq.~\ref{Rcur} into Eq.~\ref{frpoisson}, we obtain
\begin{equation}
\nabla^2f_R(R)=\frac{\Omega_m^0H_0^2}{c^2}\left(\frac{D}{f_R(R)}\right)^{\frac{1}{p_+}} - \frac{\bar{R}}{3c^2}-\frac{8\pi G\rho}{3c^2}+\frac{8\pi G\bar{\rho}}{3c^2}\quad.\label{exfrscalar}
\end{equation}
Given the density field $\rho$ and boundary conditions for $f_R$, the above equation completely determines $f_R$ on the whole simulation domain. The extra scalar field $f_R$ makes the non-linear behavior of $f(R)$ gravity very complicated. In order to better understand the impact of the extra scaler field in Eq.\ref{exfrscalar} on the large scale structure formation, we define the effective Newtonian constant as
\begin{equation}
G_{\rm eff}\equiv\left(\frac{4}{3}-\frac{\delta R}{3\kappa^2\delta \rho}\right)G,\label{Geffcon}
\end{equation}
such that the modified Poisson equation in Eq.\ref{poissonfr} can be recast into
\begin{equation}
\nabla^2\phi=4\pi G_{\rm eff}\delta \rho\quad.
\end{equation}
Clearly, $G_{\rm eff}$ directly indicates the modification of standard gravity.

In the dense regions $\rho\gg\bar{\rho}$, there are two possible types of solutions to Eq.~\ref{exfrscalar}. The gradient term on the left-hand side of Eq.~\ref{exfrscalar} can be large enough to rival the matter density field. The fact that the density is high does not mean the curvature is also very high. In this case, we have $\delta R\ll \kappa^2\delta \rho$ and the solution of Eq.~\ref{exfrscalar} is called the ``low-curvature solution''\cite{HuI}. The effective Newtonian constant $G_{\rm eff}\sim\frac{4}{3}G$ is larger than the standard gravity by a factor of $1/3$.

On the other hand, the curvature perturbation $\delta R$ can also be large enough to track the density field  $\delta R\sim\kappa^2\delta \rho$, which is known as the ``high-curvature solution'' \cite{HuI}. In this case, the modifications to standard gravity is well suppressed, and the effective Newtonian constant goes back to its GR value ($G_{\rm eff}\sim G$). If the dense regions follow the ``high-curvature solution'' at late times, the $f(R)$ model can pass local tests of gravity; this is well known as the chameleon mechanism \cite{Khoury,Mota}. However, even if at early times the dense regions generally follow the ``high-curvature solution'', at late times the solution can transfer to the ``low-curvature solution''. It is also possible that the ``high-curvature solution'' is not achieved anywhere in the universe.

At early times, the background curvature is very high ($\bar{R}\gg \bar{R}_0$ where $R_0$ is the Ricci curvature today). The density field is relatively homogenous ($\delta \rho\sim0$). The solution of Eq.~\ref{exfrscalar} is also nearly homogenous and close to the background value
\begin{equation}
f_R\sim\bar{f}_R=D\left(\frac{3\Omega_m^0H_0^2}{\bar{R}}\right)^{p_+}\quad,
\end{equation}
where
\begin{equation}\label{eq:background_R}
\bar{R}=3\Omega_m^0 H_0^2\left(\frac{1}{a^3}+\frac{4\Omega_d^0}{\Omega_m^0}\right)\quad.
\end{equation}
This is clearly the ``high-curvature solution'' since $\kappa^2\rho\sim R$.
As structure formation proceeds, $\delta R$ gradually falls behind $\kappa^2\delta\rho$ except in regions with very high $\delta\rho$, because Eq.~29 is a differential equation rather than algebraic equation. As a result, unless $\rho\gg\bar{\rho}$, we will find $G<G_{\rm eff}<4G/3$ according to Eq.~\ref{Geffcon}.

In practice, Eq.~\ref{frpoisson} is numerically solved by
using relaxation method with many iterations from the initially guessed value for the scalar field until convergence is reached. In {\sc ecosmog}, we take the initial guess for $f_R$ as its background value $\bar{f}_R$. Therefore, in dense regions where $\rho\gg \bar{\rho}$, whether we could obtain the ``high-curvature solution''  is somewhat determined by whether the value for $R$ can be efficiently boosted from $\bar{R}$ to $\kappa^2\rho$. Analytically, it can be understood like this: for given scalar curvature $R$, from Eq.~\ref{Rcur} we obtain
\begin{equation}
\tilde{\delta} R=-\frac{3\Omega_m^0 H_0^2}{p_{+}f_R}\left(\frac{D}{f_R}\right)^{\frac{1}{p_{+}}}\tilde{\delta} f_R=-\frac{R}{p_{+}f_R}\tilde{\delta} f_R\quad,\label{deltaRnum}
\end{equation}
where $\tilde{\delta}$ denotes small changes with respect to the local quantities and not the background quantities. For a given value of $R$, from Eq.~\ref{Rcur} we can see clearly that $|f_R|\rightarrow+\infty$ when $|D|\rightarrow+\infty$, which means that in Eq.~\ref{deltaRnum}, to get a small change in $R$ we need a substantial change in $f_R$. In the opposite limit, Eq.~\ref{Rcur} shows that $|f_R|\rightarrow0$ when $|D|\rightarrow0$, in which case it is easy to have significant change in $R$ with only small changes in $f_R$. Therefore, a smaller absolute value of $D$ can help form the ``high-curvature solution'' in regions where $\rho\gg \bar{\rho}$, while the larger absolute values of $D$ will do the opposite. It can then be expected that, with large $|D|$, the change $\tilde{\delta}f_R$ can be large enough for the gradient term on the left-hand side of Eq.~\ref{exfrscalar} to dominate over the curvature term on the right-hand side: in this case, $\delta R\ll\kappa^2\delta\rho$ and there is no ``high-curvature solution'' in the whole system.

After these qualitative analysis, in the next few sections, we will go through the technical details of our $N$-body simulations and present the numerical results.

\subsection{Equations in code units}

The {\sc ecosmog} code is based on the supercomoving coordinates
\begin{equation}
\begin{split}
\tilde{x}=\frac{x}{aB},\quad\rho=\frac{\rho a^3}{\rho_c\Omega_m^0},\quad \tilde{v}=\frac{av}{BH_0},\\
\tilde{\phi}=\frac{a^2\phi}{(BH_0)^2},\quad d\tilde{t}=H_0\frac{dt}{a^2},\quad \tilde{c}=\frac{c}{BH_0},
\end{split}
\end{equation}
where $x$ is the comoving coordinate, $\rho_c$ is
the critical density today, $c$ is the speed of light and $B$ is the size of the simulation box in the unit of $h^{-1}{\rm Mpc}$. In the code units,
Eq.~\ref{poissonfr} and Eq.~\ref{frpoisson} can be written as,
\begin{equation}
\tilde{\nabla}^2\tilde{\phi}=2a\Omega_m^0(\tilde{\rho}-1)
+\frac{a}{2}\Omega_m^0-\frac{a^4\Omega_m^0}{2}\left(\frac{Da^2}{\tilde{f}_R}\right)^{\frac{1}{p_+}}+2a^4\Omega_d^0,\label{codepoission}
\end{equation}
\begin{equation}
\tilde{\nabla}^2\tilde{f}_R=-\frac{a\Omega_m^0}{\tilde{c}^2}(\tilde{\rho}-1)+\frac{a^4\Omega_m^0}{\tilde{c}^2}\left(\frac{Da^2}{\tilde{f}_R}\right)^{\frac{1}{p_+}}-\frac{4a^4\Omega_d^0}{\tilde{c}^2}-\frac{a\Omega_m^0}{\tilde{c}^2},\label{codefr}
\end{equation}
where $\tilde{f}_{R}\equiv a^2 f_R$ and we have used Eq.~\ref{eq:background_R}.

Eqs.~\ref{codepoission} and \ref{codefr} here are related to the equations used in the original code for the Hu-Sawicki model \cite{HuI} by
\begin{equation}\label{model_match}
\begin{split}
n&=p_{+}-1\quad,\\
\xi&=-\frac{D}{n}3^{n+1}\quad,
\end{split}
\end{equation}
where $n$ and $\xi$ are defined in \cite{ECOSMOG}. This provides a simple way to cross-check our modification of the code. We have checked and found good agreements between our modification and the original code \cite{ECOSMOG}. For more technical issues about $N$-body simulations, the readers are referred to \cite{RAMSES,ECOSMOG}.

\subsection{Cosmological simulations}

\begin{figure*}
\includegraphics[width=7in,height=7in]{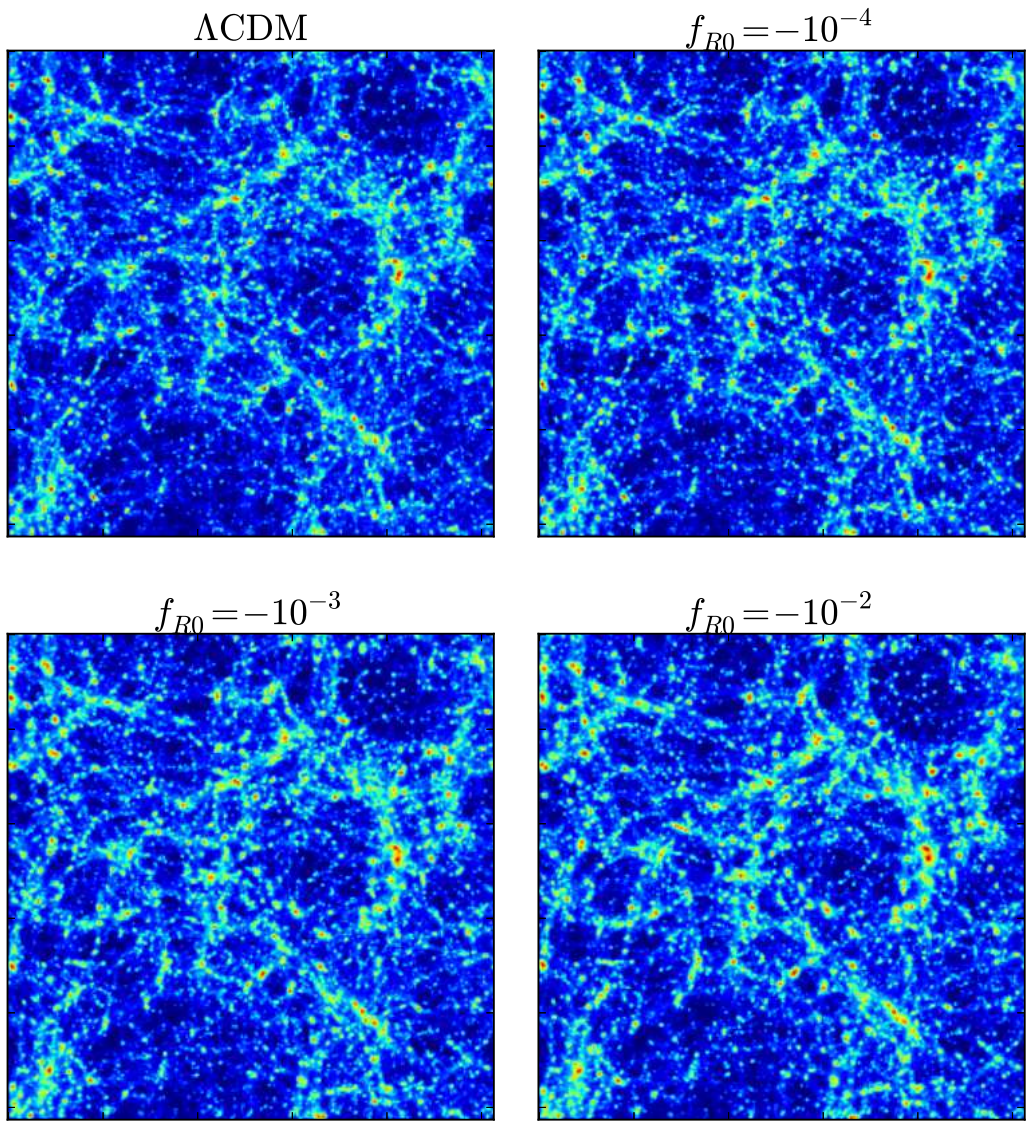}
\caption{The snapshots of density fields for $\Lambda$CDM model, $f(R)$ models with $|f_{R0}|=10^{-4},10^{-3},10^{-2}$ respectively. The snapshots are taken from the simulations with $L_{box}=150h^{-1}{\rm Mpc}$ at redshift $z=0$. }\label{visual}
\end{figure*}

In our $N$-body simulations, we adopt $\Omega_m^0=0.2814, \Omega_d^0=0.7186, h=0.697, n_s=0.962, \sigma_8=0.82$ as the cosmological parameters, which are consistent with the parameters used in the linear perturbation calculation. We use the {\sc grafic} \cite{inicon} package to generate the initial conditions, and set the starting point at $a=0.04839$, the same as in the linear calculation. In our simulations, we implement 5 realizations for each $f(R)$ model and models of the same realisation share the same initial conditions. We choose the parameter $f_{R0}$ to cover a large portion of parameter space. The detailed settings are listed in Table~\ref{settings}. In addition to the above parameters, a convergence criterion is used to determine when the relaxation method has converged. In {\sc ecosmog}, convergence is considered to be achieved when the residual of the partial differential equation, i.e., the difference between the two sides of the partial differential equation, is smaller than a predefined parameter $\epsilon$. We set $\epsilon=10^{-8}$ throughout this work. The simulation results are shown in 2D snapshots in Fig.~\ref{visual}.

In order to study the chameleon mechanism, we plot the statistics of the effective Newtonian constant $G_{\rm eff}$ with respect to the density contrast $\delta=\rho/\bar{\rho}-1$.  For this purpose, we first note down the values of the scalar field $f_{R}$ and the density field $\rho$ on the grids in the \texttt{leaves} cells (the most refined cells) that do not have \texttt{son} cells in the simulations. Then, we divide the values of $\delta$ into several bins, and count the number of cells in which the values of $\delta$ fall into each bin.  Finally, we take the arithmetical average of $G_{\rm eff}$ using Eq.~\ref{Geffcodeunit} over the cells in the simulations for each bin. Measuring $G_{\rm eff}$ provides the most straightforward way to examine the chameleon mechanism in dense regions where $\delta\gg1$,
\begin{equation}
\begin{split}
\frac{G_{\rm eff}}{G}&=\frac{4}{3}-\frac{\delta R}{3\kappa^2\delta \rho}\\
&=\frac{4}{3}-\frac{a^3}{3(\tilde{\rho}-1)}\left[\left(\frac{Da^2}{\tilde{f}_R}\right )^{\frac{1}{p_+}}-\frac{1}{a^3}-\frac{4\Omega_d^0}{\Omega_m^0} \right]\label{Geffcodeunit}
\end{split}
\end{equation}

\begin{figure}
\includegraphics[width=3.6in,height=3.2in]{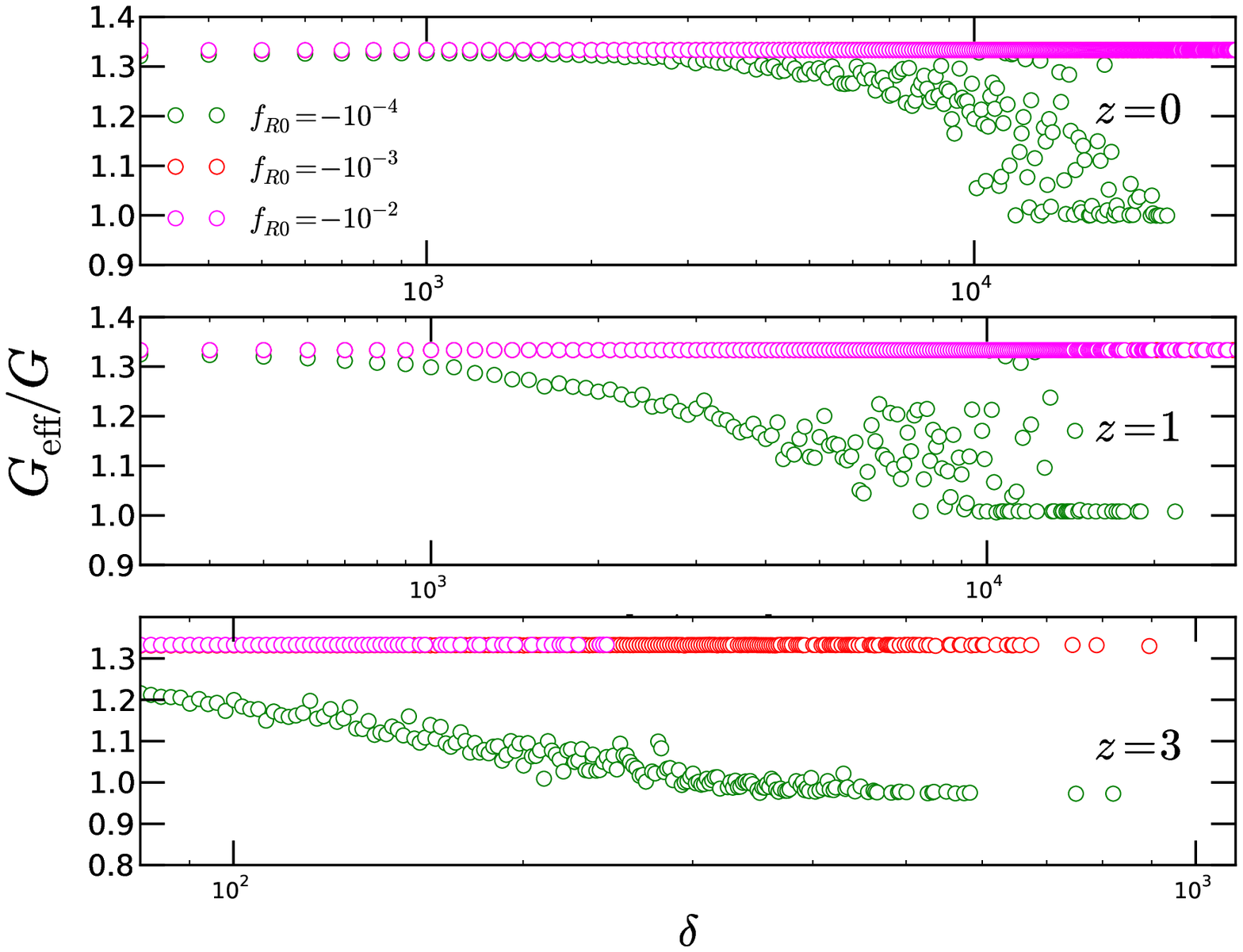}
\caption{The average effective Newtonian constant with respect to density contrast. At late time of the universe, the chameleon mechanism appears for the $f(R)$ model with $|f_{R0}|=10^{-4}$. However, for $f(R)$ models with $|f_{R0}|=10^{-2}$ and $|f_{R0}|=10^{-3}$, there are no chameleons even in the high density regions.}\label{Geff}
\end{figure}

As shown in Fig.~\ref{Geff} at late times ($z<3$), the effective Newtonian constant $G_{\rm eff}$ for $f(R)$ models with $|f_{R0}|\geq10^{-3}$ is close to $\frac{4}{3}G$ in  dense regions, which corresponds to the ``low-curvature solution'' of Eq.~\ref{frpoisson}. We find no ``high-curvature solution'' in the dense regions in these cases. On the other hand, for models with $|f_{R0}|\leq10^{-4}$, $G_{\rm eff}$ shows clear transition features from the ``high-curvature solution''  ($G_{\rm eff}\sim G$) in dense regions to the ``low-curvature solution'' ($G_{\rm eff}\sim \frac{4}{3}G$) in lower density regions. The chameleon mechanism does work, in this case, until the present time. The qualitative behavior shown by Fig.~\ref{Geff} fully agree with our previous analysis. There is an important threshold value for $|f_{R0}|$ above which we can not find ``high-curvature solution'' in the dense region in the universe at late time. Therefore, as a rough guide, viable $f(R)$ models should have $|f_{R0}|\leq10^{-4}$.

The chameleon mechanism is vital to $f(R)$ gravity not only because it can provide a way to evade the stringent  constraints from local tests of gravity, but also because it can have significant impact on the mater power spectra even on scales which are usually considered as in the linear reigme. We will explore this issue in the next subsection.

\begin{table*}
\caption{The simulation technical details about the $f(R)$ models.}\label{settings}
\begin{tabular}{c|c|c|c|c|c}
\hline
\hline
$f_{R0}$ & $B_0$ &$D$ & ${L_{box}}$ & No. of particles & realizations  \\
\hline
$-3\times 10^{-5}$ & $0.000166045$  & $-0.0000517106$& $150 h^{-1}{\rm Mpc}$& $256^3$&$5$\\
\hline
$ -5\times 10^{-5}$ & $0.000276748$   & $-0.0000861843$ & $150 h^{-1}{\rm Mpc}$&$256^3$ &$5$\\
\hline
$ -10^{-4}$ & $0.000553523$   & $-0.000172369$ & $150 h^{-1}{\rm Mpc}$&$256^3$ &$5$\\
\hline
$ -3\times10^{-4}$ & $0.0016609$   & $-0.000517106$ & $150 h^{-1}{\rm Mpc}$&$256^3$ &$5$\\
\hline
$ -10^{-3}$ & $0.00554022$   & $-0.00172369$ & $150 h^{-1}{\rm Mpc}$&$256^3$ &$5$\\
\hline
$ -5\times 10^{-3}$ & $0.0278125$   & $-0.00861843$ & $150 h^{-1}{\rm Mpc}$&$256^3$ &$5$\\
\hline
$ -10^{-2}$ & $0.0559059$   & $-0.0172369$ & $150 h^{-1}{\rm Mpc}$&$256^3$ &$5$\\
\hline
\hline
\end{tabular}\label{table}
\end{table*}
%\begin{table*}
%\begin{tabular}{cc}
%${\rm \Lambda CDM}$ & $f_{R0}=-10^{-4}$ \\
%\includegraphics[width=2.2in,height=2in]{3.eps}&\includegraphics[width=2.2in,height=2in]{4.eps}\\
%$f_{R0}=-10^{-3}$ & $f_{R0}=-10^{-2}$ \\
%\includegraphics[width=2.2in,height=2in]{5.eps}&\includegraphics[width=2.2in,height=2in]{6.eps}\\
%\end{tabular}
%\caption{The snapshots of density fields for $\Lambda$CDM model, $f(R)$ models with %$f_{R0}=-10^{-4},-10^{-3},-10^{-2}$ respectively. The snapshots are taken from the simulations %with $L_{box}=150h^{-1}{\rm Mpc}$ at redshift $z=0$.}\label{visual}
%\end{table*}

\subsection{Matter power spectra}

We use the publicly available code {\sc powmes} \cite{POWMES} to measure the matter power spectra from our simulations. {\sc powmes} constructs the density field on a regular grid by direct particle assignment and then uses fast Fourier transform to compute the spectra. The grid we used for the spectra measurement is $256^3$, which is the same as the domain grid used in our $N$-body simulations. The measured power spectra are shown in Fig.~\ref{matterPK}. In Fig.~\ref{fmatterPK}, we show the fractional differences of matter power spectra between $f(R)$ models and the $\Lambda$CDM model. The dashed lines show the predictions from linear perturbation theory, the solid lines are the linear power spectra corrected by the Halofit formula derived from $\Lambda$CDM model\cite{halofit}, and the points with error bars are measured from our simulations. Fig.~\ref{fmatterPK} shows that the non-linear power spectra have several distinct features from the results of linear perturbation theory and Halofit.

For $f(R)$ models with $|f_{R0}|\leq 10^{-4}$, Fig.~\ref{Geff} shows that the chameleon screening could be efficient from early times up until present day. The difference in the matter power spectra from the $\Lambda$CDM prediction is suppressed on all scales. The linear perturbation theory and the standard Halofit formalism can not even predict the correct qualitative behavior of the matter power spectra on small scales. Another prominent feature is that the scales of $k\sim 0.06 h{\rm Mpc}^{-1}$, which are supposed to be in the linear regime, cannot be well described by linear theory for our $f(R)$ models. Indeed, from Fig.~\ref{fmatterPK} we can see that linear theory becomes inaccurate almost as soon as the power spectrum starts to deviate from the $\Lambda$CDM prediction. The reason for this is due to the chameleon mechanism. In linear theory, the perturbation dynamics transfers from the ``high-curvature regime'' at early times to the ``low-curvature regime'' at late times, and the effective newtonian constant in all regions changes from $G$ to ${4G}/{3}$. However, in $f(R)$ simulations,  $G_{\rm eff}$ tends to be $G$ due to the chameleon mechanism both at early times and at late times in dense regions. Therefore, compared to linear theory prediction, the growth history from $N$-body simulations is closer to the $\Lambda$CDM model. In other words, the difference between $f(R)$ and $\Lambda$CDM is suppressed by the nonlinearity in the theory, as clearly shown in Fig.~\ref{fmatterPK}.

For models with $|f_{R0}|\geq10^{-3}$, the chameleon screening stops working from at least $z=3$ (see Fig.~\ref{Geff}), and the effective Newtonian constant is enhanced by ${1}/{3}$ compared to its GR value. The Halofit formalism, in such cases, can predict the matter power spectra correctly down to scales of $k\sim 0.1h^{-1}{\rm Mpc}$, because these scales are still in the linear regime with $G_{\rm eff}\sim\frac{4}{3}G$. On even smaller scales ($k>1h^{-1}{\rm Mpc}$), however, we find a significant suppression in the power spectrum. As explained in~\cite{frli}, this suppression is due to the much larger velocity dispersions at small scales, which prevent matter from even stronger clustering. Similar suppressions have been observed for non-chameleon simulations too (see Fig.9 \cite{li_barrow_2011}), and, contrary to the naive interpretation, are not because the chameleon mechanism brings things back to GR on small scales \cite{frli,Zhao,li_zhao_2010}.

\begin{figure}
\includegraphics[width=3.6in,height=3.2in]{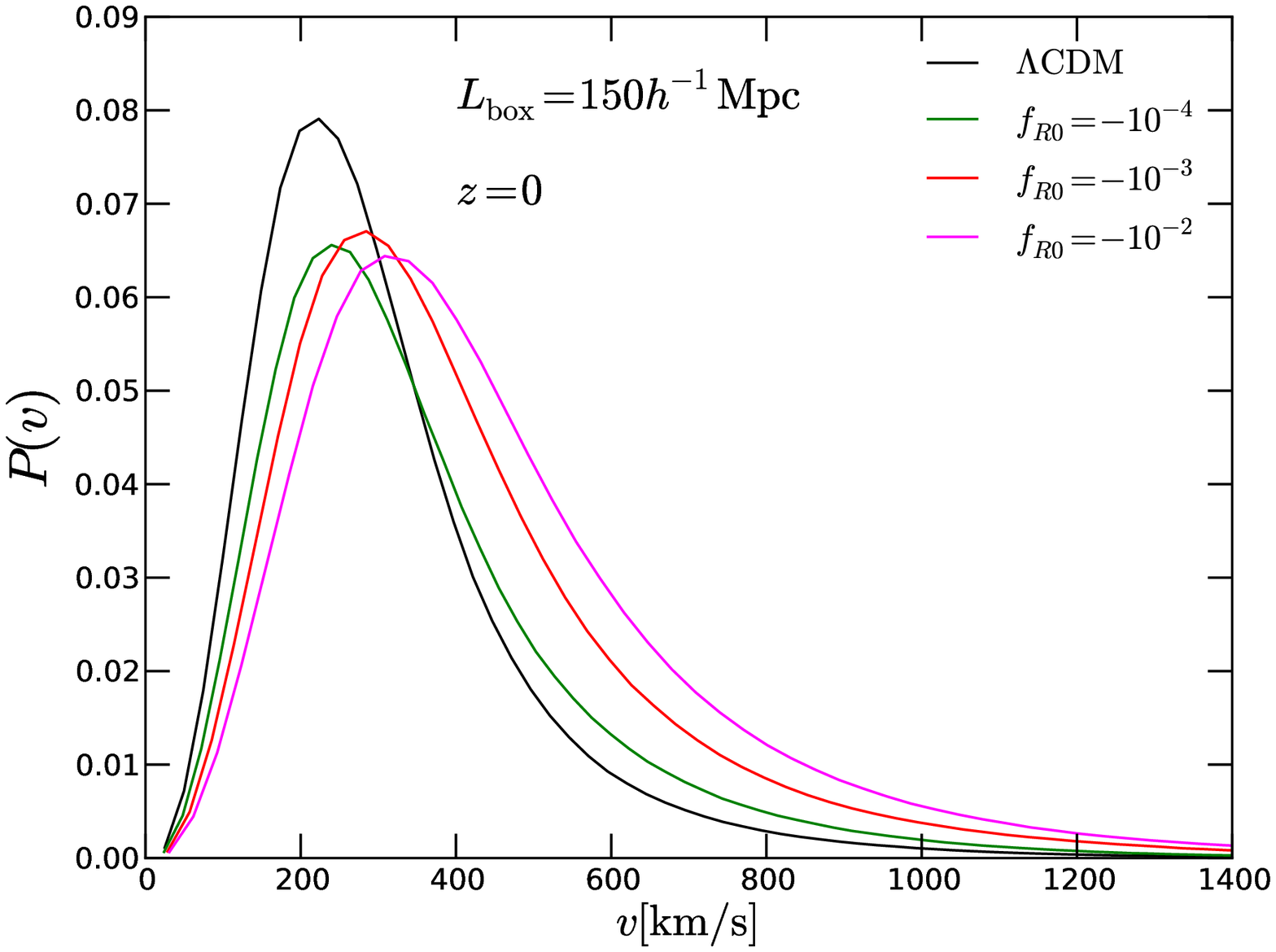}
\caption{The probability density function of particle velocity for $\Lambda$CDM model and $f(R)$ models with $|f_{R0}|=10^{-4},10^{-3},10^{-2}$ respectively.}\label{statvelf}
\end{figure}

\begin{figure}
\includegraphics[width=3.6in,height=3.2in]{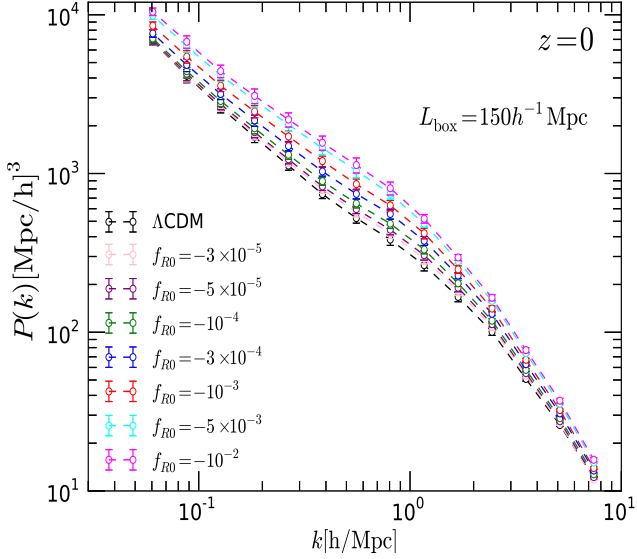}
\caption{The power spectra measured from our $N$-body simulations. The boxsize is $L_{box}=150h^{-1}{\rm Mpc}$ and the redshift is $z=0$.}\label{matterPK}
\end{figure}

\begin{figure}
\includegraphics[width=3.6in,height=3.2in]{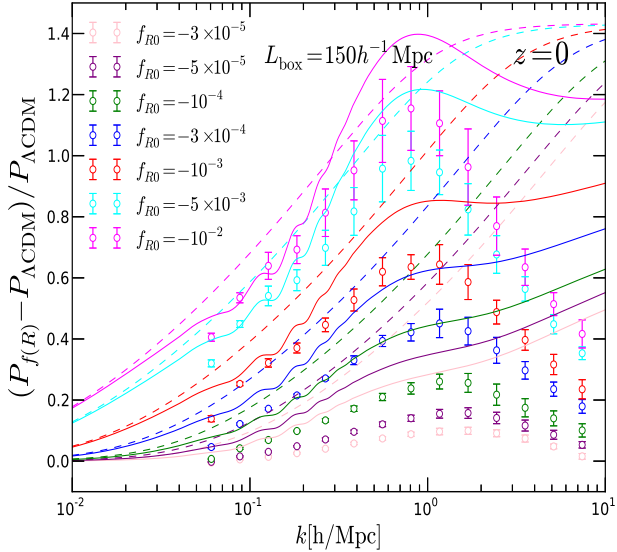}
\caption{The relative difference of the matter power
spectra between the $f(R)$ and $\Lambda$CDM simulations at z = 0. The dashed lines show the predictions from linear perturbation theory. The solid lines represent the linear power spectra corrected by the standard Halofit formula. The points with error bars are measured from our simulations.
}\label{fmatterPK}
\end{figure}

In order to quantitatively analyze the velocity dispersions, we measure the following statistical quantities
\begin{equation}
\bar{v}=\frac{1}{N}\sum_{i=1}^{N}v_i\quad,
\end{equation}
where $\bar{v}$ is the average velocity of all particles $N=256^3$ in our simulations. The velocity for each particle is defined by
\begin{equation}
v_i=\sqrt{v_{xi}^2+v_{yi}^2+v_{zi}^2}\quad.
\end{equation}
We use the standard deviation to characterize the dispersion of velocities
\begin{equation}
\sigma_v=\sqrt{\frac{1}{N-1}\sum_{i=1}^{N}(v_i-\bar{v})^2}\quad,
\end{equation}
where $\sigma_v$ has the same unit as $\bar{v}$.
In Fig.~\ref{statvelf}, we present the probability density function of particle velocity for $\Lambda$CDM model and $f(R)$ models with $|f_{R0}|=10^{-4},10^{-3},10^{-2}$ respectively. The statistical results are shown in Table~\ref{dispersion}. In the $\Lambda$CDM model, we find that the average velocity of all particles is
$\bar{v}=296.5{\rm [km/s]}$ and the dispersion is $ \sigma_v=178.4{\rm [km/s]}$. However, in $f(R)$ models we find much larger average velocity  as well as the dispersions. We find $\bar{v}=344.0{\rm [km/s]},\quad\sigma_v=210.3{\rm [km/s]}$ for model with $f_{R0}=-10^{-4}$,$\bar{v}=400.2{\rm [km/s]},\quad\sigma_v=247.0{\rm [km/s]}$ for model with $f_{R0}=-10^{-3},\quad$ and $\bar{v}=448.5{\rm [km/s]},\quad\sigma_v=272.9{\rm [km/s]}$ for $f(R)$ model with $f_{R0}=-10^{-2}$. It is clear that the larger absolute value of $f_{R0}$, the larger dispersion of the velocities in the $f(R)$ model. The increased velocity dispersion is expected to affect the profiles of halos making matter less clustered on small scales. For models with $|f_{R0}|\geq10^{-3}$, the fifth force can both accelerate particles and deepen the central potential of a halo, but particles' kinetic energy is increased more than their potential energy, so that they tend to cluster less.

\begin{table*}
\begin{tabular}{c|c|c}
\hline
\hline
Model & $\bar{v}{\rm [km/s]}$  & $\sigma_v{\rm [km/s]}$  \\
\hline
$\Lambda$CDM & $296.5 $  &  $178.4$\\
\hline
$ f_{R0}=-10^{-4}$ & $344.0 $    & $210.3 $\\
\hline
$ f_{R0}=-10^{-3}$ & $400.2$   & $247.0 $\\
\hline
$f_{R0}=-10^{-2}$ & $448.5$    & $272.9$\\
\hline
\end{tabular}
\caption{The statistical properties of the velocity field for our simulations with boxsize $L_{\rm box}=150 h^{-1}{\rm Mpc}$ at redshift $z=0$.}\label{dispersion}
\end{table*}

Compared to simulation results for the Hu-Sawicki model \cite{simulation,Zhao,frli}, for our models with no ``high-curvature solution'' at late times ($|f_{R0}|>10^{-3}$), the transition from the ``high-curvature solution'' at early times to the ``low-curvature solution'' at late times happens much earlier in the models studied here. As we shall see later in Fig.~\ref{fitpk}, the pattern of the matter power spectrum at redshift $z=3$ in our models is similar to that of the Hu-Sawicki model at $z=0$. Therefore, the qualitative behavior of our models is similar to that of the Hu-Sawicki model, but with a shift to higher redshift.

Indeed, we find that to obtain similar $P_{f(R)}/P_{\Lambda{\rm CDM}}$, the value of $|f_{R0}|$ is roughly an order of magnitude larger than the corresponding value in the Hu-Sawicki model (with $n=1$) studied in \cite{simulation,Zhao,frli}. The reason for such a difference is as follows: according to Eq.~\ref{model_match}, the models studied in \cite{simulation,Zhao,frli} corresponds to our model with $p_+=2$ while here we have $p_+\approx1.129$. A direct comparison between Eq.~\ref{codefr} above and Eq.~13 of \cite{ECOSMOG}, or Eq.~\ref{codepoission} above and Eq.~11 of \cite{ECOSMOG}, shows that the only difference is in the factor $f_R^{-1/p_+}$ (where the relationship between $D$ in our model and $\xi$ in the Hu-Sawicki model, as shown in Eq.~\ref{model_match}, is used). Clearly, as $1/p_+$ is smaller in the Hu-Sawicki model, to obtain similar $|f_R|^{1/p_+}$ (remember that $|f_R|\ll1$) their $|f_R|$ must be overall smaller.

We can also explain the observation that in our models the modified gravity effect seems to start earlier than in the Hu-Sawicki model (the shift of power spectrum pattern to higher redshift). Let us consider the background value of $|f_R|$ only, in which case we have
\begin{equation}
|\bar{f}_{R,{\rm HS}}|^{-1/(n+1)} = |\bar{f}_{R,{\rm We}}|^{-1/p_+}
\end{equation}
with $n=1, p_+=1.129$. This gives
\begin{equation}
|\bar{f}_{R,{\rm HS}}| = |\bar{f}_{R,{\rm We}}|^{2/1.129} \approx |\bar{f}_{R,{\rm We}}|^{1.77}.
\end{equation}
As $|\bar{f}_R|\ll1$ in both models, we can see $|\bar{f}_{R,{\rm We}}|\gg|\bar{f}_{R,{\rm HS}}|$ at early times. Assuming the same background cosmology for these two models, this implies that $|\bar{f}_{RR,{\rm We}}|\gg|\bar{f}_{RR,{\rm HS}}|$ and so according to Eq.~\ref{eq:compton}, the Compton wavelength would be much larger in our model at early times, resulting in an earlier effect of modified gravity. The larger Compton wavelength implies that it is much easier to violate the Compton conditions\cite{HuI} in low density regions in our model and the ``high-curvature solution'' could more easily transfer to ``low-curvature solution'' at earlier times. The large scale structure of the Universe in our model at present could be deemed as the {\it future} scenarios for Hu-Sawicki model, and our model therefore has richer phenomenology.


\subsection{Resolution issues and the PPF fit}

\begin{figure}
\includegraphics[width=3.6in,height=3.2in]{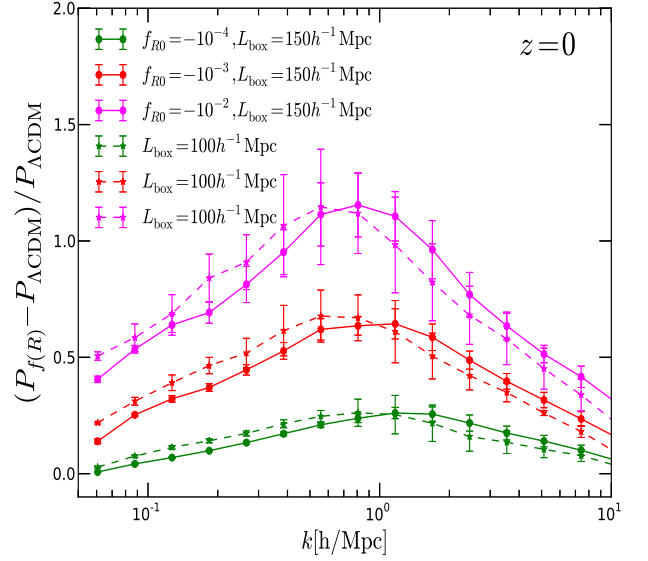}
\caption{The relative difference of the matter power
spectra between the $f(R)$ and $\Lambda$CDM simulations with different boxsize. The solid lines are for the results with $L_{box}=150h^{-1}{\rm Mpc}$ and the dashed lines for $L_{box}=100h^{-1}{\rm Mpc}$  }\label{200deltaPk}
\end{figure}

We investigate the resolution issues in $f(R)$ simulations using two different box sizes, respectively $L_{box}=150h^{-1}{\rm Mpc}$ and $L_{box}=100h^{-1}{\rm Mpc}$. To this end we choose three representative values, $|f_{R0}|=10^{-4},10^{-3},10^{-2}$, which include the $f(R)$ models both with and without chameleon screening at late times. The detailed settings are listed in Table~\ref{settings}, and the simulation results are displayed in Fig.~\ref{200deltaPk}.
On large scales ($k<1 h{\rm Mpc}^{-1}$), the simulations from the two boxes match well with each other. We find that the simulations with larger box tend to overestimate the power $\delta P/P$ on small scales $k>1 h^{-1}{\rm Mpc}$, which is consistent with what is found in \cite{frli}. Because the fifth force in $f(R)$ simulations is sensitive to the resolution, the higher-resolution simulations could give more reliable results on small scales \cite{frli}, we shall refer to the results from the smaller box hereafter.
\begin{table*}
\caption{Parameters for $f(R)$ simulations with different box sizes}\label{settings}
\begin{tabular}{c|c|c|c|c|c|c}
\hline
\hline
$f_{R0}$ & $B_0$ &$D$ & ${L_{box}}$  & ${L_{box}}$  & No. of particles & realizations  \\
\hline
$ -10^{-4}$ & $0.000553523$   & $-0.000172369$ & $100 h^{-1}{\rm Mpc}$ & $150 h^{-1}{\rm Mpc}$ &$256^3$ &$5$\\
\hline
$ -10^{-3}$ & $0.00554022$   & $-0.00172369$ & $100 h^{-1}{\rm Mpc}$& $150 h^{-1}{\rm Mpc}$&$256^3$ &$5$\\
\hline
$ -10^{-2}$ & $0.0559059$   & $-0.0172369$ & $100 h^{-1}{\rm Mpc}$& $150 h^{-1}{\rm Mpc}$&$256^3$ &$5$\\
\hline
\hline
\end{tabular}
\end{table*}

\begin{figure}
\includegraphics[width=3.6in,height=3.2in]{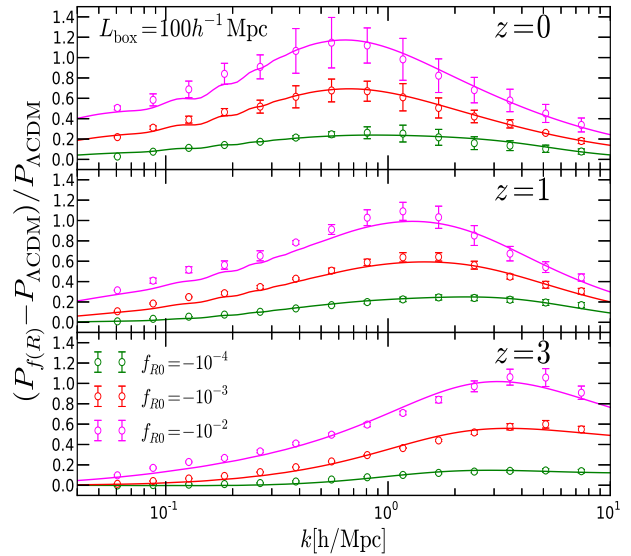}
\caption{The circles with error bars represent the results measured from simulations with  $L_{box}=100h^{-1}{\rm Mpc}$ at redshift $z=0,1,3$ respectively. The solid lines show the PPF fitting results from our generalized fitting formula.}\label{fitpk}
\end{figure}

The scale-dependent growth history of $f(R)$ gravity changes not only the amplitude but also the shape of the power spectra. In addtion, the shape of the power spectrum evolves throughout the cosmic history. To address this point, in Fig.~\ref{fitpk}, we plot $\delta P/P$ of the simulations from the $100h^{-1}{\rm Mpc}$ box at three different redshifts $z=0,1$ and $3$ respectively. The circles with error bars represent the simulation results. At redshift $z=0 $, $\delta P/P$  peaks roughly at $k\sim 0.7h{\rm Mpc}^{-1}$ for all $f(R)$ models. However, at higher redshifts, $z=1$ and $z=3$, the peaks shift to smaller scales significantly, which is roughly around $k\sim 1h{\rm Mpc}^{-1}$ at $z=1$ and $k\sim 3h{\rm Mpc}^{-1}$ at $z=3$. In \cite{frli}, such a shift is explained as the result of hierarchical structure formation: the peak position corresponding to typical cluster scales at a given time, above which matter clustering is boosted by the enhanced gravity and below which the increased velocity dispersion prevents even stronger clustering.

Hu and Sawicki has proposed a simple way to modify the Halofit to reproduce the
nonlinear power spectrum in modified gravity models, which
is called the Parameterised-Post Friedman (PPF) \cite{PPF} fit. The PPF
matter power spectrum interpolates between the non-linear power spectrum without
any screening mechanism to recover GR on small scales and the non-nonlinear power spectrum in the $\Lambda$CDM model. It assumes that on very small scales the power spectrum should go back to the $\Lambda$CDM result, and a simple form is given by \cite{PPF}
\begin{equation}
P(k,z)=\frac{P_{\rm non-GR}(k,z)+c_{\rm nl}\Sigma^2(k,z)P_{\rm GR}(k,z)}{1+c_{\rm nl}\Sigma^2(k,z)}\quad,\label{fittingEq}
\end{equation}
where $P_{\rm non-GR}$ indicates the non-linear power spectrum in modified gravity without
the mechanism that recovers GR on small scales, and in our case can be simply taken as the linear power spectrum in $f(R)$ gravity corrected by the standard Halofit formula.

$P_{\rm GR}$ is the power spectrum in $\Lambda$CDM model. $\Sigma^2(k,z)$ is given by
\begin{equation}
\Sigma^2(k,z)=\left[\frac{k^3}{2\pi^2}P_{\rm lin}(k,z)\right]^{1/3}\quad.
\end{equation}
$P_{\rm lin}$ is the linear power spectrum in $f(R)$ gravity. Eq.~\ref{fittingEq} has been tested and shown to work very well in several modified gravity models \cite{Zhao,frli}.  However, we find that this simple formula gives poor fits to our simulations by overestimating the power on small scales $k>1h{\rm Mpc}^{-1}$. In order to get a better fitting, we generalize Eq.~\ref{fittingEq} by making the coefficient of $c_{\rm nl}$ as a function of $k$:
\begin{equation}
P(k,z)=\frac{P_{\rm non-GR}(k,z)+(C_{\rm nl1} k^{\alpha} +C_{\rm nl2})\Sigma^2(k,z)P_{\rm GR}(k,z)}{1+(C_{\rm nl1} k^{\alpha} +C_{\rm nl2})\Sigma^2(k,z)},
\end{equation}
in which $C_{\rm nl1}$ and $C_{\rm nl2}$ are dimensionless fitting parameters which depend on model and redshift, and so is $\alpha$.

The performance of our modified fitting formula are shown in Fig.~\ref{fitpk} as solid lines. The best-fit PPF parameters are listed in table \ref{fitting}. Although the generalized fitting formula works very well for individual models, it is still challenging to find a single formula which could fit well for all these models at different redshifts. The reason is twofold. First, the growth history is scale dependent, and the shape of the power spectrum varies with redshift. Second, the chameleon mechanism works for models with $|f_{R0}|<10^{-4}$ but not for models with $|f_{R0}|>10^{-3}$: it is hard to mediate the formula from the models with chameleon mechanism to those without.

\begin{table*}
\caption{The best-fit PPF parameters}\label{fitting}
\begin{tabular}{c|ccc|ccc|ccc}
\hline
\hline
Redshift & &$z=0$ & & &$z=1$ & & &$z=3$ &\\
$f_{R0}$ & $-10^{-4}$ &$-10^{-3}$ & $-10^{-2}$  & $-10^{-4}$ &$-10^{-3}$ & $-10^{-2}$ & $-10^{-4}$ &$-10^{-3}$ & $-10^{-2}$ \\
\hline
$ C_{\rm nl1}$ & $0.02349462$   & $0.1410763$ & $0.1212703$ & $0.02247135$ &$0.05641741$ &$0.05899864$ & $0.3860476$ &$0.01189381$ &$0.05077320$\\
$ C_{\rm nl2}$ & $0.4634951$   & $0.01632510$ & $0.01721348$ & $0.1484467$ &$0.003195103$ &$0.03894015$ & $0.3163662$ &$0.1535029$ &$0.01491219$\\
$\alpha$ & $2.251794$   & $1.129913$ & $1.036022$ & $1.990064$ &$1.426073$ &$1.296817$ & $0.4359099$ &$0.6882835$ &$0.3786083$\\
\hline
\hline
\end{tabular}
\end{table*}

\section{conclusions\label{conclusions}}

In this work, we have studied the impact of a family of $f(R)$ models that can reproduce the $\Lambda$CDM background expansion history on the large-scale structure using a large suite of $N$-body simulations. We have analyzed the chameleon mechanism using our simulation data, and found that it works throughout the whole cosmic history (in dense regions) provided that $|f_{R0}|<10^{-4}$ in our model. However, for models with $|f_{R0}|>=10^{-3}$, we find no ``high-curvature solution'' in dense regions at late times (e.g., $z<3$), which means that those models could be ruled out due to the factor-of-${1}/{3}$ enhancement to the strength of Newtonian gravity. Although our simulations have limited resolution, our results do show that the chameleon mechanism fails to bring the value of $|f_R|$ to be very small inside dark matter halos for models with $|f_{R0}|>=10^{-3}$. There is no thin-shell structures observed in these simulations. The galaxies's gravitational potentials are not sufficient to make them self-screened, and as the galaxies are not screened, the stars' potentials are not sufficient to make them self-screened either. As the screening mechanism fails for both galaxies and stars, the model can be safely ruled out.

We have analyzed the non-linear matter power spectra for our $f(R)$ models. Compared to simulation results for the Hu-Sawicki model \cite{simulation,Zhao,frli}, our models show much smaller deviations from $\Lambda$CDM for the same value of $|f_{R0}|$, as is shown clearly in the plot of $\delta P/P$; equivalently, to get the same deviation from the $\Lambda$CDM power spectrum, our model requires larger values of $|f_{R0}|$. The modified gravity effect starts earlier in our models than in the Hu-Sawicki model, and this can be explained by the difference in the values of the parameter $p_+$ in these two models.

We have also generalized the PPF fitting formula \cite{PPF} to fit our simulation results, and the new fitting formula works very well for individual $f(R)$ models. However,
it is still challenging to find a single formula which could fit well for all these models at different redshifts, due to the scale-dependent growth history and the chameleon effect.

Finally, it is very important to note that even in the model with $|f_{R0}|<10^{-4}$  where the  chameleon mechanism could work efficiently in the dense regions and there are no significant signatures in the matter power spectra, in low density regions where $\rho\sim \bar{\rho}$ or in voids where $\rho \sim 0$, the Compton condition \cite{HuI} is violated and the strength of the gravity could substantially differ from the GR result, which provides a smoking gun for testing the modified gravity theories, as pointed out by \cite{voids}. It is therefore very interesting to investigate the halo and void properties in our $f(R)$ model, and this will be a subject for future work.

\emph{Acknowledgment: J.H.He acknowledges the Financial
support of MIUR through PRIN 2008 and ASI through
contract Euclid-NIS I/039/10/0. BL is supported by the Royal Astronomical Society
and Durham University. YPJ is sponsored by NSFC  (11121062, 11033006) and the CAS/SAFEA International Partnership
Program for  Creative Research Teams  (KJCX2-YW-T23).
 }


\begin{thebibliography}{99}

\bibitem{1} S.~J.~Perlmutter {\it et~al.},  Nature {\bf391}, 51 (1998); A.~G.~Riess {\it et~al.}, Astron.~J., {\bf116}, 1109 (1998); S.~J.~Perlmutter {\it et~al.}, Astrophys.~J., {\bf517}, 565 (1999); J.~L. Tonry {\it et~al.}, Astrophys.~J., {\bf594}, 1 (2003); A.~G.~Riess {\it et~al.}, Astrophys.~J., {\bf607}, 665 (2005); P.~Astier {\it et~al.}, Astron.~Astrophys., {\bf447}, 31 (2006); A.~G.~Riess {\it et~al.}, Astrophys.~J., {\bf659}, 98 (2007).

\bibitem{WMAP} E.~Komatsu {\it et.~al.}, Astrophys.~J.~Suppl., {\bf192}, 18 (2011); P.~A.~R.~Ade {\it et.~al.} (2013), arXiv:1303.5076.

\bibitem{BAOm} A.~G.~Sanchez {\it et.~al.} (2012), arXiv:1203.6616.

\bibitem{sean} S.~M.~Carroll, Living~Rev.~Rel., {\bf4}, 1 (2001).

\bibitem{fr} P.~G.~Bergmann, Int.~J.~Theor.~Phys., {\bf1}, 25 (1968); A.~A.~Starobinsky, Phys.~Lett.~B{\bf91}, 99 (1980); A.~L.~Erickcek, T.~L.~Smith and M.~Kamionkowski, Phys.~Rev.~D{\bf74}, 121501 (2006); V.~Faraoni,  Phys.~Rev.~D{\bf74}, 023529 (2006); S.~Capozziello and S.~Tsujikawa, Phys.~Rev.~D{\bf77}, 107501 (2008); T.~Chiba, T.~L.~Smith and A.~L.~Erickcek,  Phys.~Rev.~D{\bf75}, 124014 (2007); I.~Navarro  and K.~Van~Acoleyen, J.~Cosmo.~Astropart.~Phys., {\bf02}, 022 (2007); G.~J.~Olmo, Phys.~Rev.~Lett., {\bf95}, 261102 (2005); G.~J.~Olmo,  Phys.~Rev.~D{\bf72}, 083505 (2005); L.~Amendola, D.~Polarski and S.~Tsujikawa, Phys.~Rev.~Lett., {\bf98}, 131302 (2007); L.~Amendola, R.~Gannouji, D.~Polarski and S.~Tsujikawa, Phys.~Rev.~D{\bf75}, 083504 (2007); L.~Amendola, Phys.~Rev.~D{\bf60}, 043501 (1999).

\bibitem{solution} T.~Multamaki and I.~Vilja, Phys.~Rev.~D{\bf73}, 024018 (2006); S.~Nojiri and S.~D.~Odintsov, Phys.~Rev.~D{\bf74}, 086005 (2006); S.~Nojiri and S.~D.~Odintsov, J.~Phys.~A{\bf40}, 6725 (2007);  S.~Capozziello, S.~Nojiri, S.~D.~Odintsov and A.~Troisi, Phys.~Lett.~B{\bf639}, 135 (2006); K.~Bamba, C.-Q.~Geng, S.~Nojiri and S.~D.~Odintsov, Phys.~Rev.~D{\bf79}, 083014 (2009); S.~Carloni, R.~Goswami and P.~K.~S.~Dunsby (2010), arXiv:1005.1840; R.~Myrzakulov, D.~Saez-Gomez and A.~Tureanu, Gen.~Rel.~Grav., {\bf43} 1671 (2011).

\bibitem{Brans} C.~Brans and R.~H.~Dicke, Phys.~Rev., {\bf124}, 925 (1961).

\bibitem{Dicke} R.~H.~Dicke, Phys.~Rev., {\bf125}, 2163 (1962).

\bibitem{Hefr}  J.-h.~He, B.~Wang and E.~Abdalla, Phys.~Rev.~D{\bf84}, 123526 (2011).

\bibitem{Maeda} K.-I.~Maeda, Phys.~Rev.~D{\bf39}, 3159 (1989).

\bibitem{conT} S.~Carloni, E.~Elizalde and S.~Odintsov, Gen.~Rel.~Grav., {\bf42}, 1667 (2010).

\bibitem{Magnano}  G.~Magnano and L.~M.~Sokolowski, Phys.~Rev.~D{\bf50}, 5039 (1994)

\bibitem{Fujii} Y.~Fujii, Prog.~Theor.~Phys., {\bf118}, 983 (2007).

\bibitem{frmodel} J.-h.~He and B.~Wang, Phys.~Rev.~D{\bf87}, 023508 (2013).

\bibitem{CAMB} A. ~Lewis, A. ~Challinor and A. Lasenby, Astrophys. ~J{\bf 538} 473 (2000).

\bibitem{frlinear} J.-h.~He, Phys.~Rev.~D{\bf86}, 103505 (2012).

\bibitem{Mota} D.~F.~Mota and J.~D.~Barrow, Phys.~Lett.~B{\bf581}, 141 (2004).

\bibitem{Khoury} J.~Khoury and A.~Weltman, Phys.~Rev.~D{\bf69}, 044026 (2004); J.~Khoury and A.~Weltman, Phys.~Rev.~Lett., {\bf93}, 171104 (2004).

\bibitem{ECOSMOG} B.~Li, G.-B.~Zhao, R.~Teyssier and K.~Koyama, J.~Cosmo.~Astropart.~Phys., {\bf1}, 51 (2012).

\bibitem{frreview} A.~Silvestri and M.~Trodden, Rept.~Prog.~Phys., {\bf72}, 096901 (2009); T.~Clifton, P.~G.~Ferreira, A.~Padilla and C.~Skordis, Phys.~Rept.~{\bf 513},1 (2012); T.~P.~Sotiriou and V.~Faraoni, Rev.~Mod.~Phys.,~{\bf82}, 451 (2010).

\bibitem{review_Tsujikawa} A.~De~Felice and S.~Tsujikawa, Living.~Rev.~Rel., {\bf13}, 3 (2010);

\bibitem{Dombriz} A.~de~la~Cruz-Dombriz, A.~Dobado, A.~L.~Maroto, Phys.~Rev.~D{\bf 77}, 123515 (2008).

\bibitem{li_barrow_2007} B.~Li and J.~D.~Barrow, Phys.~Rev.~D{\bf75}, 084010 (2007).

\bibitem{HuI}  W.~Hu and I.~Sawicki, Phys.~Rev.~D{\bf76}, 064004 (2007).

\bibitem{brax_2008} P.~Brax, C.~van~de~Bruck, A.~C.~Davis and D.~J.~Shaw, Phys.~Rev.~D{\bf78}, 104021 (2008).

\bibitem{Song}  Y.-S.~Song, W.~Hu and I.~Sawicki, Phys.~Rev.~D{\bf75}, 044004 (2007).

\bibitem{RAMSES} R.~Teyssier, Astron.~\&~Astrophys., {\bf385}, 337 (2002).

\bibitem{inicon} E.~Bertschinger (1995), arXiv:astro-ph/9506070.

\bibitem{POWMES} S.~Colombi, A.~H.~Jaffe, D.~Novikov and C.~Pichon, Mon.~Not.~R.~Astron.~Soc., {\bf393}, 511 (2009).

\bibitem{halofit} R.~E.~Smith {\it et al.} Mon.~Not.~R.~Astron.~Soc., {\bf341} 1311 (2003).

\bibitem{simulationpk} H.~Oyaizu, M.~Lima and W.~Hu, Phys.~Rev.~D{\bf78},123524 (2008).

\bibitem{frli} B.~Li, W.~A.~Hellwing, K.~Koyama, G.-B.~Zhao, E.~Jennings and C.~M.~Baugh (2012), arXiv:1206.4317.

\bibitem{PPF} W.~Hu and I.~Sawicki, Phys.~Rev.~D{\bf76}, 104043 (2007).

\bibitem{li_barrow_2011} B.~Li and J.~D.~Barrow, Phys.~Rev.~D{\bf83}, 024007 (2011).

\bibitem{Zhao} G.-B.~Zhao, B.~Li and K.~Koyama, Phys.~Rev.~D{\bf83}, 044007 (2011).

\bibitem{li_zhao_2010} B.~Li and H.~Zhao, Phys.~Rev.~D{\bf81}, 104007 (2010).

\bibitem{simulation} H.~Oyaizu, Phys.~Rev.~D{\bf78}, 123523 (2008); F.~Schmidt, M.~V.~Lima, H.~Oyaizu and W.~Hu (2008), arXiv:0812.0545.

\bibitem{voids} G.-B.~Zhao, B.~Li and K.~Koyama, Phys.~Rev.~Lett., {\bf107}, 071303 (2011);
                           B.~Li, G.-B.~Zhao and K.~Koyama, Mon.~Not.~R.~Astron.~Soc., {\bf421}, 3481 (2012).
\end{thebibliography}
\end{document}